\documentclass[ %
 reprint,
 amsmath,amssymb,
 aps,
]{revtex4-2}
\usepackage{xcolor} % for \textcolor
\usepackage{graphicx}% Include figure files
\usepackage{dcolumn}% Align table columns on decimal point
\usepackage{bm}% bold math
\begin{document}

\preprint{APS/123-QED}

\title{Highly Polarized and Long Range Dissipationless Spin Transport Due to Counterflowing Electron and Hole Edge Channels}% Force line breaks with \\

\author{M. Cosset-Ch\'eneau}
\author{B. Yang}%
\author{B.J. van Wees}%
% \email{Second.Author@institution.edu}

\affiliation{%
 Zernike Institute for Advanced Materials, University of Groningen, 9747 AG Groningen, The Netherlands
}%

\begin{abstract}
The presence of edge channels in the quantum Hall regime leads to dissipationless charge transport over long distances. When graphene is interfaced with a magnetic material, the exchange interaction lifts the Landau levels spin degeneracy. This causes the presence of counterflowing edge channels with opposite spin polarization. We show theoretically that the spin-flip scattering between these edge channels enables a dissipationless spin transport with larger than 100\% spin polarization of the charge current. It also allows the transport of spin over macroscopically long distances, even in the absence of an applied charge current.
\end{abstract}

%\keywords{Suggested keywords}%Use showkeys class option if keyword
                              %display desired
\maketitle

%\tableofcontents

Efficiently injecting, transporting and detecting a spin current while dissipating a minimal amount of energy is essential for the development of spintronics, both in terms of well-established magnetic memory devices \cite{ZuticRMP2004, ManchonRMP2019, YangNature2022}, but also for more recent device concepts based on magnon spin injection-based devices \cite{CornelissenNP2015, LebrunNature2018, WalPRB2023}, and for the development of 2D light sources with potential for long range transport of spin information \cite{ZuticSSC2020, CarrascoAPL2025}. Spin current injection mechanisms rely either on the flow of a charge current from a ferromagnet into a neighboring material \cite{JedemaPRB2003,HanNN2014}, or on spin-charge interconversion effects \cite{SinovaRMP2015,TrierNRM2022}. The efficiency of these mechanisms remains however limited, as indicated by the low spin-charge interconversion coefficients \cite{GalceranAPLM2021} and by the smaller than 100\% spin polarization of the charge current in ferromagnets \cite{ZahndPRB2018}. This is caused by the fact that, in these mechanisms, each electron flowing across the system injects at most one spin [$\hbar/2$] angular momentum. In addition, the spin current can only be transported over limited distances \cite{Bass2007}. The development of spintronics platforms in which each electron could inject more than one spin into the system, and where the spin could be transported over long distances, could therefore expand the range of phenomena accessible by spintronics experiments.

\begin{figure}[ht]
    \centering
    \includegraphics[scale=1]{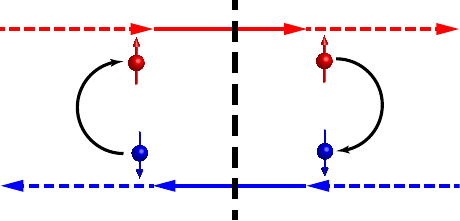}
    \caption{Schematics of the electron spin-flip scattering between counterflowing spin up (red) and down (blue) channels at one edge of a magnetic graphene ribbon in the QHE regime. An electron can cross the section represented by the dashed line with opposite spins while propagating in opposite direction, effectively contributing to more than one spin passage to the spin current.}
    \label{fig:fig1}
\end{figure}

In conventional ferromagnets, this smaller than one spin polarization of the charge current is caused by the absence of locking between the electronic spin and momentum. The flow of minority spins partially compensates the flow of majority spins \cite{ValetPRB1993}. A spin-momentum locking can be found in topological phases such as in the magnetic Quantum Hall Effect (QHE) \cite{YoungNature2014} and in the Quantum spin Hall effects (QSHE) \cite{KonigScience2007,BruneNP2012} in which the edge channels topological protection enables the dissipationless transport of spin information over long distances \cite{KonigPRX2013}. These edge channels have been used to perform spin injection resulting in magnon excitation and detection \cite{StepanovNP2018,WeiScience2018}.  In the QSHE, the phase coherence must however be maintained, and its loss intrinsically limits the distance over which the spin can be transported.

The recent observations \cite{YangNC2024,ShinAFM2024} of spin-polarized counterflowing edge channels in magnetic graphene at relatively high temperatures with a gate tunable spin-polarization \cite{YangNC2024} lifts this limitation. Magnetic graphene is a system in which a graphene layer is interfaced with an insulating magnetic material that lifts the spin degeneracy in the graphene band structure by an exchange shift \cite{GhiasiNN2021}. In the quantum Hall phase, this allows electron and hole Landau levels (LLs) with opposite spin states to be simultaneously populated when the Fermi level is placed between the spin up and down zeroth LLs \cite{YangNC2024}. Here, electron and hole LLs refers to their upward or downward energy dispersion at the edges. The system therefore hosts counterflowing edge channels with opposite spin polarizations, which means that the carrier spin polarization are locked with their momentum. However, in contrast to the QSHE, the edge channels in magnetic graphene are topologically protected against scattering from one edge to the other, but not against spin-flip scattering.

\begin{figure}[ht]
    \centering
    \includegraphics[scale=0.5]{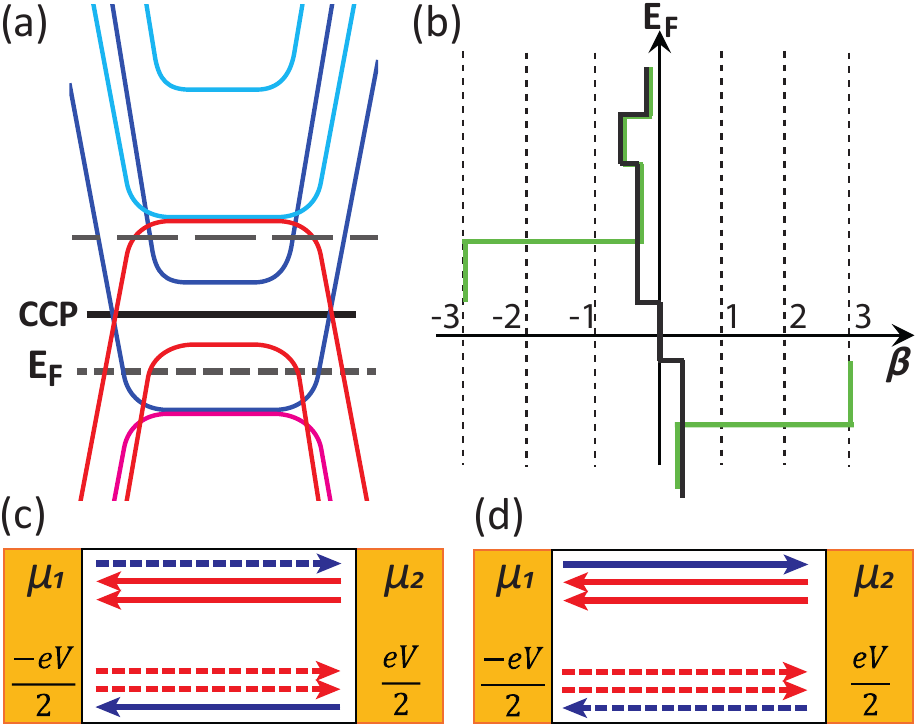}
    \caption{(a) Energy dispersion of the spin-split Landau levels in magnetized graphene. The spin up electron (hole) Landau levels are plotted in light blue (red) and the spin down electron (hole) Landau levels are in dark blue (pink). They produce electron and hole like spin polarized counterflowing edge channels. When the Fermi energy $E_F$ is between the spin up and down zero Landau levels, the spin up and down carriers flow in opposite directions. (b) Spin polarization $\beta$ of the charge current as a function of the Fermi energy for the non-equilibrated (black) and fully equilibrated (green) case. Edge transport for the non-equilibration (c) and full equilibration (d) of the counterflowing edge channels. In (c), the dashed (solid) channel electrochemical potential is $\mu_1$ ($\mu_2$). In (d), the dashed (solid) channels are at electrochemical potentials $\mu_b$ ($\mu_t$).}
    \label{fig:fig2}
\end{figure}

In this paper, we show that, in magnetic graphene, the spin-flip scattering enables a long range non-dissipative spin transport. We discuss theoretically the  effect of the spin-flip induced counterflowing edge channel equilibration on the charge and spin transport in the quantum Hall phase of this system. We show that the equilibration between the spin-polarized edge channels enables a dissipationless spin current over long distance, with a spin polarization of the charge current larger than 100 \%. In addition, a spin accumulation at one contact can generate a spin current which transports spin over macroscopic distances.

In magnetic graphene, a carrier flowing in a given direction can experience  spin-flip scattering, which also corresponds to a backscattering event [Fig. 1]. This causes an equilibration of the counterflowing edge channel electrochemical potentials between the contacts \cite{AkihoPRB2019}, and therefore a relaxation of the spin accumulation. Due to the spin-momentum locking, the spin relaxation length $\lambda_0$ is proportional to $v_F \tau$ \cite{ZhangPRB2016}  with $v_F$ the carrier velocity and $\tau$ its scattering time. All the interchannel scattering effects are included in $\lambda_0$. This sets an equilibration length $\lambda\propto \lambda_0$ of the electrochemical potential. For a device length $L\ll \lambda$, the equilibration is negligible and a ballistic spin current is present at the edges. For $L\gg \lambda$ the electrochemical potentials are fully equilibrated, and the LLs with opposite spin polarization are at the same electrochemical potentials. Note that here we assume that the edge channels with the same spin and same type equilibrate over a distance much smaller than $\lambda_0$ \cite{WeesPRL1989}. The longer equilibration length for electron and hole edge channels is caused by their opposite spin polarization, as backscattering requires the flipping of the carrier spin.

We first discuss a situation in which magnetic graphene is contacted with two metallic reservoirs (contacts 1 and 2) submitted to a voltage bias $V=V_2-V_1$. The contact electrochemical potentials are $\mu_1=-eV/2$ and $\mu_2=eV/2$ [Figs. 2(c) and 2(d)].
The Fermi energy is placed below the charge compensation point [CCP in Fig. 2(a)] such that $N^\uparrow$ spin up hole (flowing counterclockwise) and $N^\downarrow$ spin down electron (flowing clockwise) edge channels are occupied. Here, $N^{\uparrow(\downarrow)}$ includes the LLs valley degeneracy.

Equilibration between counterflowing edge channels is initially assumed to be absent [Fig. 2(c)].
The charge current normalized by $(e/h)$ is $I_c=(N^\uparrow+N^\downarrow ) eV$, while the spin current normalized by $(1/4 \pi)$ writes $I_s=(N^\uparrow-N^\downarrow ) eV$.  The spin polarization $\beta=I_s/I_c$ of the charge current is 
\begin{equation}
\beta=\frac{N^\uparrow-N^\downarrow}{N^\uparrow+N^\downarrow}.    
\end{equation}
Its range is $-1<\beta<1$, similar to the charge current spin polarization in a ferromagnetic metal.

However, in the presence of a strong equilibration between counterflowing edge channels at the same edge [Fig. 2(d)], all channels flowing at the top edge are populated up to $\mu_t$ while at the bottom edge they are populated up to $\mu_b$. In this case, $I_c=(N^\uparrow-N^\downarrow) (\mu_b-\mu_t)$ and $I_s=(N^\uparrow+N^\downarrow ) (\mu_b-\mu_t)$ such that the polarization is 
\begin{equation}
    \beta=\frac{N^\uparrow+N^\downarrow}{N^\uparrow-N^\downarrow},
\end{equation}
which is valid for $N^\uparrow \neq N^\downarrow$. This corresponds to a unique situation in which $|\beta|\geq 1$. This larger than 100\% polarization originates from the locking between spin and flow direction, causing all the electrons flowing in the edge channels to contribute to the spin current regardless of their propagation direction [Fig. 1]. In addition, $\beta$ does not depend on the spin relaxation length $\lambda_0$ when it is small enough to achieve full equilibration (see below). A reduction of the spin relaxation length therefore induces an increase of $\beta$ until it saturates. Similar to the case of charge transport in the QHE regime, the spin transport is not affected by additional spin relaxation effects or by the presence of contacts, and corresponds to a long range dissipationless spin current as long as there is no scattering between opposites edges.

In Fig. 2(b) we show $\beta$ as a function of the Fermi level position at $T=0$. Importantly, this result remains valid as long as the thermal energy ($k_bT$) is smaller than the LL spacing. When the Fermi energy is located slightly above or below the bottom of a LL, the spin polarization can even largely exceed the values predicted by eq. (2) (see supplemental material \cite{SM}). 

In contrast to the case of the charge transport in the QHE regime, a dissipation region will be present close to the contacts due to spin dissipation, in addition to inside the contacts. In the following, we study the evolution of the electrochemical potentials along the entire device length to assess the effect of dissipation close to the contacts on the spin transport, and to evaluate in which conditions a larger than one polarization can be reached.

In the rest of the discussion, we consider a situation in which all the electron (hole) edge channels are spin down (up) polarized with occupation number $N^\downarrow$ ($N^\uparrow$). For simplicity we assume that $N^\uparrow > N^\downarrow$. The electrochemical potential of the spin up and down LLs on the top ($t$) edge are denoted $\mu_t^\downarrow$ and 
$\mu_t^\uparrow$, flowing in the right and left directions, respectively. They are denoted $\mu_b^\downarrow$ and $\mu_b^\uparrow$ on the bottom edge ($b$), flowing in the left and right directions.

The normalized spin current at the top and bottom edges are $I_s^t=-N^\uparrow \mu_t^\uparrow-N^\downarrow \mu_t^\downarrow$ and $I_s^b=N^\uparrow \mu_b^\uparrow+N^\downarrow \mu_b^\downarrow$ which gives for $N^\uparrow\neq N^\downarrow$: 
\begin{equation}
    I_s^{t(b)} =\epsilon_{t(b)}  (N^\uparrow-N^\downarrow)(\mu_s^{t(b)} +\beta \mu_m^{t(b)} ),
\end{equation}
With $\epsilon_b=-\epsilon_t=1$. We note $\mu_m^{t(b)}=(\mu_{t(b)}^\uparrow+\mu_{t(b)}^\downarrow)/2$ the average electrochemical potential of the opposite spin species, and $\mu_s^{t(b)}=(\mu_{t(b)}^\uparrow-\mu_{t(b)}^\downarrow)/2$ the spin accumulation. This shows that the presence of non-fully equilibrated counterflowing edge channels implies a position dependent spin accumulation which contributes to the spin current. For fully equilibrated edges channels ($\mu_s^{t(b)}=0$), the spin current scales with the charge current as $j_s^{t(b)}= \beta j_c^{t(b)}$, and is therefore dissipationless.

The spin relaxation causes an electron transfer $j_{\mathrm{trans}}^t = (1/\lambda_0)(\mu_t^\downarrow - \mu_t^\uparrow)$
 and $j_{\mathrm{trans}}^b =(1/\lambda_0)(\mu_b^\uparrow - \mu_b^\downarrow)$ (normalized current) per unit of length between the counterflowing channels at a given edge.  Due to the directionality of the edge transport in the QHE regime, the electrochemical potentials follow \cite{SM}

\begin{equation}
    \frac{\mathrm{d}\mu_t^\sigma}{\mathrm{d}x}=\frac{1}{\lambda_0 N^\sigma}(\mu_{t}^\uparrow - \mu_{t}^\downarrow)
\end{equation}
\begin{equation}
    \frac{\mathrm{d}\mu_b^\sigma}{\mathrm{d}x}=\frac{1}{\lambda_0 N^\sigma}(\mu_{b}^\downarrow - \mu_{b}^\uparrow)
\end{equation}
with $\sigma=\uparrow$ or $\downarrow$. The spin accumulations at the top and bottom edges are therefore solutions of  $\partial_x \mu_s^b = -\mu_s^b/\lambda$ and $\partial_x \mu_s^t = \mu_s^t/\lambda$ with $\lambda$ the edge channel spin equilibration length

\begin{equation}
    \lambda = \lambda_0 \left( \frac{1}{N^\downarrow} - \frac{1}{N^\uparrow} \right)^{-1},
\end{equation}
the sign of which depends on the spin of the majority carriers ($N^\uparrow$ or $N^\downarrow$). The general form of the electrochemical potentials is $\mu_t^\sigma(x) = (\alpha/N^\sigma)\exp(-x/\lambda) + \gamma_t$ and $\mu_b^\sigma(x) = (\beta/N^\sigma)\exp(x/\lambda) + \gamma_b$ for $N^\uparrow\neq N^\downarrow$. 

This is illustrated for $N^\uparrow=3$ and $N^\downarrow=2$  in Figs 3(a).  Assuming ideal contacts, the edge channels flowing away from contact $i$ are fully occupied up to $\mu_i$ and the channels incident on the contact are fully absorbed. The spin up (down) channels propagate anticlockwise (clockwise). The boundary conditions are $\mu_t^\downarrow(0)=\mu_b^\uparrow(0)=\mu_1$ and $\mu_t^\uparrow(L)=\mu_b^\downarrow(L)=\mu_2$. The computed electrochemical potentials are plotted in Fig. 3(c). 
\begin{figure}[ht]
    \centering
    \includegraphics[scale=1]{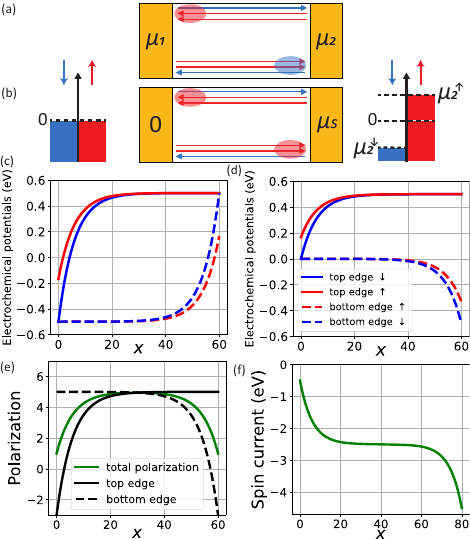}
    \caption{Schematics of the device under consideration, with $N^\uparrow=3$ and $N^\downarrow=2$ when an electrochemical potential bias $\mu$ is applied between the contacts (a), or a spin accumulation bias $ \mu_s$ is applied at contact 2 (b). (c) Spatial dependence of the edge channels electrochemical potentials in (a). The x-position is measured in units of $\lambda_0$. See inset in (d) for the legend. (d) Edge channels electrochemical potential for (b). (e) Spatial dependence of the spin polarization of the charge current flowing at the top edge (solid black lines), bottom edge (black dashed) and of the total spin current (green) for (a). (f) Spin current induced by the application of the spin accumulation bias in (b).}
    \label{fig:fig3}
\end{figure}

For $L\ll \lambda$, (not shown), equilibration does not take place and the electrochemical potentials are constant, equal to the one at the contact from which the carriers are emitted [Fig. 2(c)]. When $L \gg \lambda$ [Fig. 3(c)], the electrochemical potentials evolve along the edges. This evolution depends on the relative number of spin up and down edge channels, with the electrochemical potential of the minority edge channels converging toward the electrochemical potential of the majority ones. From this picture, we conclude that in Fig. 2(d) $\mu_t=\mu_2$ and $\mu_b=\mu_1$.  At the top edge, $\mu_t^\uparrow>\mu_t^\downarrow$ and a positive spin accumulation is present in the vicinity of contact 1. This spin accumulation is opposite at the bottom edge close to contact 2.  These spin accumulation areas are spin "hotspots" [blue and green-shaded area in Fig. 3(a)], in which the transport becomes dissipative. For $N^\downarrow>N^\uparrow$, the spin hotspot at the top (bottom) edges is in the vicinity of contact 1 (2). We show in Ref. \cite{SM} how these hotspots can be electrically detected using ferromagnetic contacts.

In Fig. 3(e) we show the polarization (spin current normalized by the charge current) at both edges. We also compute the net spin current normalized by the charge current flowing between the contacts. These quantities correspond to the spin polarization of the charge current $\beta$ introduced above. The polarizations at the top and bottom edges (green lines) are not equal and have different signs close to the contacts. Note however that similar to the charge current in the QHE regime, the only relevant quantities for transport measurements are the difference of spin and charge current at the top and bottom edges, which corresponds to measurable net spin and charge currents. In principle, and similar to the charge transport, a circulating spin current may be present in magnetic graphene. The way to detect this current however remains unclear.

The spin polarization of the net charge current [green line in Fig. 3(e)] is larger than 1 away from the contacts, following the above discussion. It however drops to 1 at the interface with contacts 1 and 2. This value of the spin polarization remains constant for $L\gg \lambda$ if both electron and hole edge channels are occupied, independently of the specific value of $\lambda$.
The spin current emitted by one contact is therefore equal to the one received by the other contact. Magnetic graphene is a perfect spin transmitter in the two carrier regime. This is confirmed by Landauer-Buttiker computation \cite{SM} showing that, in the two carrier regime with $L\gg \lambda$ and $N^\uparrow\neq N^\downarrow$, the charge current flowing from contact 1 to contact 2 is $I_c=[N^\uparrow-N^\downarrow]\mu$, equal to the spin current emitted from contact 1 and received by contact 2. The case $N^\uparrow=N^\downarrow$ is discussed in Ref. \cite{SM}.

The difference in spin polarization at the contacts and away from them indicates that the spin hotspots correspond to regions where a spin current $I_{s,\mathrm{inj}}^{t(b)}$ is injected into or extracted from the environment by mechanisms such as (in)elastic spin-flip scattering, or transferred to other spin systems such as a magnon bath by magnon annihilation and generation. From the conservation of the spin angular momentum, these spin currents writes $  I_{s,\mathrm{inj}}^{t(b)}=2\int \frac{1}{\lambda_0} \left( \mu^\uparrow_{t(b)}-\mu^\downarrow_{t(b)} \right) \mathrm{d}x$.

This yields  $I_{s,\mathrm{inj}}^{t}=N^\downarrow \mu$ and $I_{s,\mathrm{inj}}^{b}=-N^\downarrow \mu$, meaning that the spin injected at the top channel hotspot are recovered at the bottom channel hotspot. In addition, $| I_{s,\mathrm{inj}}^{t(b)}/I_c|=N^\downarrow/(N^\uparrow - N^\downarrow)$, illustrating that each electron flowing across the system can inject (or extract) more than one spin to or from a given hotspot. In Fig. 3(a), the red shaded area corresponds to a region in which the spins are injected into the environment, while they are extracted from it in the blue area. Note that the position of these hotspots depends on whether $N^\uparrow>N^\downarrow$ or $N^\uparrow<N^\downarrow$.

The power dissipated at each hotspot region is $P_{\mathrm{hp}}=\int j_{\mathrm{trans}}^{t(b)}\left(\mu_{t(b)}^{\uparrow(\downarrow)}-\mu_{t(b)}^{\downarrow(\uparrow)} \right) \mathrm{d}x$ which in the fully equilibrated case and assuming $N^\uparrow>N^\downarrow$ yields $P_{\mathrm{hp}}=I_c^2 N^\downarrow/[2N^\uparrow(N^\uparrow-N^\downarrow)]$ for an imposed bias current $I_c$. The dissipation in each contact is $P_{\mathrm{cont}}=I_c^2/(2 N^\uparrow)$. The total dissipated power is therefore $P_{\mathrm{tot}} = I_c^2 /(N^\uparrow - N^\downarrow)$, equal to the power supplied by the current source. An increase of spin injection at the hotspots has therefore an energetic cost due to the decrease of the conductance with $N^\uparrow - N^\downarrow$.

A long range spin current can also be induced in this system by applying a spin accumulation bias in one of the contacts without applying a charge current. We consider a situation where $\mu_1=\mu_2=0$. A spin accumulation is imposed in contact 2 [Fig. 3(b)], such that $\mu_t^\uparrow (L)= \mu_s$ and $\mu_t^\downarrow (L)=- \mu_s$ with $ \mu_s = \left( \mu_2^\uparrow - \mu_2^\downarrow \right)/2 $ the spin accumulation at contact 2. The  boundary conditions at contact 1 are $\mu_t^\downarrow(0)=\mu_b^\uparrow(0)=0$. Here, two spin hotspots are induced in magnetized graphene in the vicinity of the contacts [Fig. 3(d)]. In contrast to the case described in Fig. 3(a), however, the sign of the spin accumulation is the same at both hotspots. A positive spin accumulation has therefore been transported from one contact to the other. The spin accumulation bias also induces a charge current, $I_c= \mu_s (N^\downarrow - N^\uparrow)$ for $L\gg \lambda$, which corresponds to a spin-charge interconversion effect. Finally, for $L\gg \lambda$, the spin currents at contact 2 are $I_s (L)=  -\mu_s (N^\uparrow + 3 N^\downarrow)$ and $I_s(0)= \mu_s (N^\downarrow - N^\uparrow)$ at contact 1. They are independent from $L$, which means that a spin current is transported from contact 2 to contact 1 over a distance which can be macroscopically long in the absence of scattering between edges. The spin current emitted from contact 2 which does not reach contact 1 has been transferred to the environment at the hotspots.

In conclusion, in graphene interfaced with a magnetic material, the spin-flip scattering induced equilibration between counterflowing spin polarized edge channels results in a long range dissipationless spin current with larger than 100\% spin polarization away from the contacts. The dissipation takes place near the contacts, resulting in the presence of spin hotspots. At the contacts, the spin polarization is 100\%, showing a perfect transmission of spin current from one contact to the other. These spin hotspots correspond to regions where a flow of angular momentum is transferred to the magnetic material with which graphene is interfaced, offering a possibility for gate tunable magnon injection and a long range transport of spin accumulation even in the absence of an applied charge current.

This work was supported by Zernike Institute for Advanced Materials (ZIAM), the Spinoza prize awarded to B.J.v.W. by the Nederlandse Organisatie voorWetenschappelijk Onderzoek (NWO) in 2016, and has received funding from the European Union under the ERC Advanced Grant 2DMAGSPIN (Grant agreement No. 101053054). The authors acknowledge the research program “Materials for the Quantum Age” (QuMat) for financial support. This program (registration number 024.005.006) is part of the Gravitation program financed by the Dutch Ministry of Education, Culture and Science (OCW).

\end{document}